\documentclass[12pt]{article}
\usepackage[dvips]{graphicx}
\usepackage{amssymb}
\usepackage{amsmath}
\usepackage{epsfig}

\usepackage{epsf}
\usepackage{graphicx,epsfig}
\usepackage{amsfonts}
\usepackage{amssymb}

\def\del{\partial}

\def\gsim{\, \rlap{$>$}{\lower 1.1ex\hbox{$\sim$}}\,}
\def\lsim{\, \rlap{$<$}{\lower 1.1ex\hbox{$\sim$}}\,}

\makeatletter
\renewcommand\section{\@startsection {section}{1}{\z@}%
                                 {-3.5ex \@plus -1ex \@minus -.2ex}
                                   {2.3ex \@plus.2ex}%
                                   {\normalfont\large\bfseries}}
\renewcommand\subsection{\@startsection{subsection}{2}{\z@}%
                                   {-3.25ex\@plus -1ex \@minus -.2ex}%
                                     {1.5ex \@plus .2ex}%
                                     {\normalfont\bfseries}}
\renewcommand\subsubsection{\@startsection{subsubsection}{3}{\z@}%
                                   {-3.25ex\@plus -1ex \@minus -.2ex}%
                                     {1.5ex \@plus .2ex}%
                                     {\normalfont\itshape}}
\makeatother

\newcommand{\be}{\begin{equation}}
\newcommand{\ee}{\end{equation}}
\newcommand{\bea}{\begin{eqnarray}}
\newcommand{\eea}{\end{eqnarray}}
\newcommand{\barr}{\begin{array}}
\newcommand{\earr}{\end{array}}

\def\beq{\begin{equation}}
\def\eeq{\end{equation}}
\def\be{\begin{equation}}
\def\ee{\end{equation}}
\def\bea{\begin{eqnarray}}
\def\eea{\end{eqnarray}}

\DeclareRobustCommand{\SkipTocEntry}[4]{}

\begin{document}

\begin{titlepage}

\setcounter{page}{1} \baselineskip=15.5pt \thispagestyle{empty}

\begin{flushright}
NSF-KITP-09-151\\
SU-ITP-09/38\\
SLAC-PUB-13748
\end{flushright}
\vfil

\begin{center}
{\LARGE  Dual Purpose Landscaping Tools:\\  Small Extra Dimensions in AdS/CFT
}

\end{center}
\bigskip\

\begin{center}
{\large Joseph Polchinski$^1$ and Eva Silverstein$^{1,2}$}
\end{center}

\begin{center}
\textit{$^1$Kavli Institute for Theoretical Physics and Department of Physics, University of California, Santa Barbara CA 93106
}
\end{center}

\begin{center}
\textit{
$^2$SLAC and Department of Physics, Stanford University, Stanford CA 94305
}
\end{center} \vfil

\noindent

We propose a class of AdS/CFT dual pairs which have small internal dimensions on the gravity side.  Starting from known Freund-Rubin AdS/CFT dual pairs, we use 7-branes to nearly cancel the curvature energy of the internal dimensions while maintaining their stabilization.
This leads to a new corner of the landscape -- a class of AdS solutions with a hierarchically large AdS radius -- with a dual field theory given (implicitly) by the infrared limit of a concrete brane construction involving D3-branes, 7-branes, and curvature. We first construct a class of hierarchical AdS5/CFT4 dual pairs with a simple formula for the number of degrees of freedom which we interpret in the dual QFT.  We then generalize these to AdS4/CFT3 duals, and suggest extensions of the method to obtain de Sitter solutions.

\vfil
\begin{flushleft}
\today
\end{flushleft}

\end{titlepage}

\newpage
\tableofcontents
\newpage

\section{Introduction and Motivation}
\label{sec:intro}

A long-term goal of research in quantum gravity is to formulate four-dimensional physics non-perturbatively.  Although this ultimately requires cosmology, a more proximate goal ~\cite{Landdualattempts,Banks:2003es} is the construction of $AdS_4/CFT_3$ dual pairs~\cite{Maldacena:1997re} with some of the most basic requirements of realism, such as ${\cal N}\le 1$ supersymmetry and small internal dimensions -- {\it i.e.} a gap between the masses of four-dimensional particles and those of internal degrees of freedom such as Kaluza-Klein modes.\footnote{There are interesting earlier approaches to four-dimensional quantum gravity in string theory using Matrix Theory \cite{Banks:1996vh}\ or the $AdS_2\times S^2$ version of AdS/CFT.  These may work but are subject to significant infrared problems.}  Indeed, many constructions of accelerating and inflating vacua in string theory can be viewed as ``uplifting" AdS vacua, adding additional ingredients with positive potential energy.  Moreover, the AdS case and its generalizations figure in potential applications of string theory to the study of strongly correlated systems; in these applications, formulating a landscape of examples based on compactification rather than consistent truncation\footnote{We thank A. Dabholkar for this concise characterization.} is also of interest.

Even this is challenging.  The Freund-Rubin spaces obtained from the simplest near-horizon limits of branes a la Ref.~\cite{Maldacena:1997re}\ have enormous internal dimensions, and do not admit uplifting to de Sitter space.
This occurs because the curvature of the internal dimensions balances against that of the AdS dimensions in the solution.
The more general $AdS_4$ landscape vacua that have been constructed \cite{Landscapereviews}\ are not directly related to any known brane constructions.  The weak curvature of spacetime means that the gravity side is the effective, weakly coupled description (if any) of the system, and the dual involves strongly coupled, non-supersymmetric quantum field theory which is difficult to derive.

In this paper, we propose a solution to this problem and illustrate it with a new class of compactifications with small internal dimensions.  Our strategy is to begin with a known AdS/CFT dual pair, obtain small internal dimensions by adding ingredients which on the gravity side nearly cancel the curvature of the internal dimensions, and interpret the result on the field theory side.
The additional ingredients we use are (p,q) 7-branes of type IIB string theory, which we analyze in detail using F theory \cite{Vafa:1996xn,MorrisonVafa}. On the gravity side, these 7-branes -- corresponding to a $T^2$ fibration in the F theory language -- contribute potential energy of the same order and opposite sign to that descending from the positive curvature of the internal base compactification manifold.  This suggests a method for constructing $AdS\times Small$ solutions with hierarchically large AdS radius by tuning a discrete parameter to be large in such a way that the 7-branes nearly but incompletely cancel the curvature energy.

To illustrate this, we present explicit brane constructions consisting of D3-branes in F theory placed at the tip of narrow, noncompact Calabi-Yau four-fold cones.  Implementing our construction requires understanding the physical status of geometric singularities in these Calabi-Yau manifolds.  We suggest a criterion for physically allowed singularities, apply it in our examples, and discuss further subtleties arising at the tip of the cone in a class of explicit models.

The dual field theory for a given solution is given by the low energy limit of its brane construction.
Aside from this implicit definition, we will not determine the dual field theories in any detail in this paper.  However, the construction unveils several characteristic features of the relevant field theories.  Most interestingly,
the (p,q) 7-branes presumably correspond to electric and magnetic flavors \cite{threeseven}.  This suggests that the relevant CFTs involve a generalization of Argyres-Douglas fixed points~\cite{Argyres:1995jj} to field theories with less supersymmetry.  Secondly, the solutions allow us to determine the number of degrees of freedom (central charge) of the dual field theory.  The small internal dimensions yield an enhanced central charge relative to the underlying Freund-Rubin example, and the corresponding narrow shape of the cone allows us to interpret the parametric dependence of this number of degrees of freedom on the data of the Calabi-Yau construction in terms of field theoretic degrees of freedom.

We will start with $AdS_5\times Small$ examples involving D3-branes, 7-branes, and geometry and discuss subtle issues to do with singularities on these spaces.  Then we will generalize these to $AdS_4\times Small$ solutions in two ways.  We will discuss future directions and potential generalizations, including methods for obtaining de Sitter solutions.
It will be interesting to see if our results can help illuminate the problem of formulating inflating backgrounds non-perturbatively.

Although we have been led to a new corner of the landscape in the present construction, this class has some key ingredients -- such as 7-branes and fluxes -- in common with previous constructions in the landscape such as \cite{GKP,Kachru:2003aw,Balasubramanian:2005zx,Denef:2005mm,Saltman:2004jh}.  Our examples share the simple feature of \cite{DeWolfe:2005uu}\ of having a parametric power-law hierarchy.
It will be interesting to see if the present methods ultimately extend to some of these cases.

\section{General Considerations}
\label{sec:general}
\setcounter{equation}{0}

\subsection{Hierarchies in AdS/CFT}
\label{subsec:hierarchies}

Our goal is to find $AdS_d/CFT_{d-1}$ duals\footnote{$d=4$ being the case of most interest.  We will also denote the dimension of the CFT by $d' = d-1$.} where the AdS radius
$R_{AdS} $ is much larger than the compactification radius $R_{comp}$ of the remaining directions.
Since the string scale $\ell_s$ cannot be larger than the compactification scale, and the $d$-dimensional Planck scale $\ell_P$ cannot be larger than the string scale, we have as our goal
\begin{equation}
R_{AdS} \gg R_{comp} \gsim \ell_s \gsim \ell_P\ .
\end{equation}

By considering the entropy of an AdS black hole~\cite{Witten:1998zw} we can conclude that the effective number of degrees of freedom
\beq\label{Ndof} N_{d.o.f.}\sim R_{AdS}^{d-2} /l_P^{d-2} \eeq
must be large if there is a hierarchy.  The AdS/CFT relation for the $AdS_{d}$ mass~\cite{Gubser:1998bc,Witten:1998qj},
\begin{equation}
m^2 R_{AdS}^2 = \Delta(\Delta - d+1) \ ,
\end{equation}
implies that the operators dual to Kaluza-Klein modes have must have large dimensions.  In contrast to the known examples, where there are large numbers of protected operators dual to the KK modes, only a small number of low energy states may retain small dimensions.  Any weakly coupled field theory will have many operators with dimensions of order one, which must become parametrically large when we have a hierarchy.  Thus a necessary condition is that the coupling must be strong.

The large number of degrees of freedom and the strong coupling are no surprise, but one can also draw less obvious conclusions about the amount of supersymmetry.   In many cases, if there is an $R$-symmetry group, it will protect a large number of operators and lead to a large compact dimension.  For example, the $d'=4$, ${\cal N}=4$ gauge theory has an $SO(6)$ $R$-symmetry and the protected operators are dual to the Kaluza-Klein states on a large $S^5$.  For $d'=4$, ${\cal N} = 2$ and $d'=3$, ${\cal N}=4$ the $R$-symmetries are $SU(2)\times U(1)$ and $SU(2)\times SU(2)$, suggesting a large $S^2 \times S^1$, $S^2 \times S^2$ or $S^3$.  For $d'=4$, ${\cal N} = 1$ and $d'=3$, ${\cal N}=2$ the $R$-symmetry is $U(1)$, suggesting a large $S^1$.\footnote{The large $S^1$'s can be reduced in size by orbifolding on a $Z_N$, which restricts the $R$-charge to multiples of $N$ and breaks the supersymmetry.  A similar effect can also occur even with the supersymmetry unbroken, as we will see.}  With $d'= 3$, ${\cal N}= 1$ supersymmetry in the CFT, leading to $d = 4$, ${\cal N}= 1$ supersymmetry in the bulk (counting the doubling due to superconformal invariance), there is no $R$-symmetry and no protected operators.

One can illustrate the role of the $R$-symmetry with some familiar examples.  For the $AdS_3 \times S^3 \times T^4$ duals, the $S^3$ radius must be the same as the AdS radius, but the size of the $T^4$ can be much smaller.  In this case, the $R$ symmetry acts on the $S^3$ coordinates but not on the $T^4$ coordinates.  In the IIA examples studied in Ref.~\cite{DeWolfe:2005uu}, there is a large hierarchy and indeed the supersymmetry is ${\cal N}=1$; unfortunately a CFT dual is still unknown.

This discussion suggests the natural conjecture that sufficient conditions for the desired large hierarchy would consist of a large number of degrees of freedom, strong coupling, and ${\cal N}\leq 1$ supersymmetry, on the grounds that with strong coupling and no $R$-symmetry essentially all operators will acquire large anomalous dimensions.  However, we have not been able to find examples realizing this simple strategy.  In many examples operator dimensions appear to be protected due to inheritance from more supersymmetric theories, as in orbifolding~\cite{Kachru:1998ys}.  In others there are anomalous dimensions that are large in the sense of being of order one, but not parametrically large.

\subsection{Curvature and Seven-Branes}
\label{subsec:curvature}

In Freund-Rubin compactifications, the internal dimensions live on a positively curved Einstein manifold ${\cal Y}$.  In the solution to Einstein's equations, the curvature of ${\cal Y}$ balances against the AdS curvature $\sim 1/R_{AdS}^2$ and against stress-energy from flux.  These three contributions are all of the same order in the solution, so the curvature radius of ${\cal Y}$ is of order $R_{AdS}$.

It will be useful to describe this equivalently in terms of the effective theory in the AdS directions.
Consider compactifying string theory (or M theory) down to $d$ dimensions on a manifold ${\cal Y}$ of dimensions $D-d$.  Among the various contributions to the potential energy for scalar fields in the remaining dimensions \cite{Landscapereviews}\ is a contribution ${\cal U}_{\cal R}$ obtained from dimensionally reducing the higher-dimensional Einstein action $\int \frac{d^Dx}{\ell_P^{D-2}}\sqrt{G}{\cal R}$ (where $\ell_P$ is the $D$-dimensional Planck length).  Let us focus on string-theoretic models in which $\ell_P^{D-2}=g_s^2\alpha^{'(D-2)/2}$ where $g_s$ is the string coupling and $\alpha'$ the inverse string tension, and further specialize to $D=10$.
We will mostly focus on the cases $d=5$ and $d=4$.

In $d$-dimensional Einstein frame, the potential energy descending from curvature is of order
\beq\label{curvaturepot} {\cal U}_{\cal R}\sim -M_d^d \left(\frac{g_s^2}{Vol_{\cal Y}}\right)^{\frac{2}{d-2}}\frac{1}{R_{\cal Y}^2}  \eeq
where $M_d$ is the $d$-dimensional Planck mass, $Vol_{\cal Y}$ is the volume of ${\cal Y}$, and $R_{\cal Y}$ is the curvature radius of ${\cal Y}$.  That is, we have taken ${\cal Y}$ to have positive curvature ${\cal R}_{\cal Y}\sim 1/R_{\cal Y}^2$.\footnote{The ten-dimensional Einstein equations include constraints; in general one must ensure that the geometry -- combined with fluxes and other ingredients -- gives consistent initial data in GR.}
The exponent ${2}/({d-2})$ arises as $-1$ from the string frame effective action and $+{d}/({d-2})$
from the Weyl transformation of the effective potential.

For example, the $AdS_5\times S^5$ solution of type IIB string theory with $N_c$ units of 5-form RR flux arises along the minimum of the potential
\beq\label{FRSfive} {\cal U}_{{\cal R}+F}\sim M_5^5 \left(\frac{g_s^2}{Vol_{S^5 }}\right)^{\frac{2}{3}}\left(-\frac{1}{R_{S^5}^2}+ \frac{g_s^2 N_c^2}{Vol_{S^5}^2} \right)\ .
\eeq
The last factor comes from the internal curvature and five-form flux terms in the string frame action, with numerical constants set to one.   This description in terms of an effective $d$ dimensional action requires a consistent truncation since the internal Kaluza-Klein modes are not separated in scale from the light modes on AdS; it is a useful method for estimating scales even when exact solutions are not available.  Extremizing with respect to $R_{S^5}$ one finds that the two terms are of the same order,\footnote{In general one would also have to extremize with respect to the dilaton $g_s$, but in this case there is a marginal direction and this is redundant.} and therefore also of the same order as the AdS curvature term in the potential.  It is possible to use orbifolds by some discrete group $\Gamma$ to reduce the size of ${\cal Y}$ below its curvature radius in a subset of the directions \cite{Kachru:1998ys}, but this procedure still leaves some directions in which ${\cal Y}/\Gamma$ is as large as $R_{AdS}$.

The same scaling holds for general $AdS_5/CFT_4$ Freund-Rubin compactifications supported by 5-form flux, which arise as gravity duals of the infrared limit of D3-branes at the tip of non-compact Calabi-Yau threefolds which are cones over Einstein spaces.  The Freund-Rubin relation between the AdS radius and the internal radii, $R_{AdS}\sim R_{\cal Y}$ corresponds to the fact that the angular distance around the cone is of order the distance to its tip.  There has been extensive work developing both sides of the duality for large classes of examples of this sort~\cite{Ypq,Labc,Hananyetal}.

In order to avoid this conclusion, one needs some offsetting term in the potential.  The possibilities are limited, because the curvature term tends to dominate at large radius and weak coupling and so drive the vacuum energy negative.  However, in type IIB string theory in 10 dimensions, stress-energy from 7-branes competes with curvature energy.  There are several ways to see this.  First, somewhat loosely speaking, since 7-branes are at real codimension two, their contribution to the stress tensor scales like $1/R^2$ times a hard cosmological constant, just like curvature.  Of course, as real codimension-two objects, 7-branes back react strongly on the geometry.  The effect of this is properly accounted for by F theory, in which the varying axio-dilaton $\tau=C_0+i/g_s$ sourced by the 7-branes corresponds to the complex structure of a $T^2$ fibered over space \cite{Vafa:1996xn}.

For example, in the eight-dimensional compactification of F theory on an elliptically fibered Calabi-Yau manifold, the 24 7-branes exactly cancel the positive curvature of the ${\cal Y} = S^2$ base manifold, leaving the noncompact dimensions flat.  Thus, our  basic idea is to consider examples where ${\cal Y} $ corresponds to a Freund-Rubin compactification, and a set of 7-branes -- equivalently the elliptic fibration of F theory -- cancels or nearly cancels its curvature energy at fixed internal size.  This reduced curvature balances against the AdS curvature, yielding a larger AdS radius.  Such a cancellation might be parametric, as in Ref.~\cite{DeWolfe:2005uu}, or sporadic, as for example in Ref.~\cite{Kachru:2003aw}.  In the framework that we use it will be natural to look for a parametric cancellation.

Consider the naive potential energy from 7-branes (which we will use F theory to study reliably below).  Each 7-brane fills the $d$ noncompact spacetime dimensions and wraps a codimension-2 $(8-d)$-dimensional cycle $\Sigma$ of volume $Vol_\Sigma$ in ${\cal Y}$.  The potential energy for a 7-brane of tension $\tau_7$ naively scales as
\beq\label{Useven} {\cal U}_7\sim \left(\frac{g_s^2}{Vol_{\cal Y}}\right)^{\frac{d}{d-2}}\tau_7 Vol_\Sigma
= \left(\frac{g_s^2}{Vol_{\cal Y}}\right)^{\frac{2}{d-2}}\tau_7 g_s^2\left(\frac{Vol_\Sigma}{Vol_{\cal Y}}\right) \eeq
The last factor scales like $(Length)^{-2}$ as does curvature, and comparing to (\ref{curvaturepot}) we see that there is therefore a potential for 7-branes to cancel some or all of the curvature energy.  At this level, such a cancellation requires more than just D7-branes; we need (p,q) 7-branes of tension $\tau_7\propto 1/g_s^2$ in order to match the factors of $g_s$.
Although this was heuristic, this conclusion will hold in the appropriate F theory description of 7-branes.
With our additional 7-branes, we will be led to quantum field theories arising as the infrared limit of D3-branes in F theory on noncompact elliptically fibered Calabi-Yau four-folds.  In order to obtain a hierarchy, we require that all of the angular directions be smaller than the radial distance to the tip.  In the next section we will implement this explicitly.

Besides the potential cancellation, the introduction of 7-branes seems promising from another point of view.  The starting point for many landscape constructions is a Ricci-flat manifold, or more generally a manifold with negative scalar curvature.  So roughly speaking our goal is to turn a sphere into a Calabi-Yau, to go from the Einstein spaces that are present in the known AdS/CFT duals to the Ricci-flat (or negatively curved) spaces that form the starting point for many landscape constructions.  F theory \cite{Vafa:1996xn}\ provides such a connection.


Now let us discuss the general features of the near-horizon geometry.
The simplest examples of the AdS/CFT correspondence are $AdS_5/CFT_4$ dual pairs obtained by a compactification on a 5-dimensional Einstein manifold.  On the other hand, F-theory is most easily formulated on a complex base manifold.  For this reason -- and also because the following structure will arise naturally in our ultimate noncompact brane construction -- we will consider 5-manifolds ${\cal Y}_5$ which are $S^1$ fibrations over a K\"ahler base manifold ${\cal B}$ of complex dimension two:
\begin{eqnarray} \label{hopf} S^1_f \rightarrow  &{\cal Y}_5& \nonumber \\
&\downarrow & \nonumber \\ &{\cal B}&
\end{eqnarray}
When the $S^1$ is small -- as we will find in our solutions -- the compactification on ${\cal Y}_5$ can be regarded as a compactification on ${\cal B}$ with metric flux (i.e. gauge flux of the Kaluza-Klein $U(1)$ descending from the $S^1$ fiber).  We will use (\ref{hopf}) for compactifications to $d=5$, and later generalize to $d=4$.

In the next section, we will show using F theory how to arrange 7-branes in such a way as to nearly cancel the curvature energy of ${\cal B}$ with a small relative factor of $\epsilon$ related to discrete quantum numbers we will introduce.  We will also consider combinations of 7-branes at which the string coupling is extremized at order one.

Given this, the five-dimensional effective potential relevant for the $d=5$ case contains the terms
\beq\label{stabmech} {\cal U}\sim M_5^5 (R_f R^4)^{-2/3}
\left(\frac{R_f^2}{
R^4}-\frac{\epsilon}{
R^2}+\frac{N_c^2}{R^8 R_f^2}\right) \eeq
where $R_f\sqrt{\alpha'}$ is the size of the fiber circle $S^1_f$, and $R\sqrt{\alpha'}$ is the size of ${\cal B}$.  The first term is from the metric flux of the $S^1$ fiber, and the second is the net contribution of the internal curvature and seven-branes, reduced by $\epsilon$ due to the near cancellation.
In this F theory setting there is no global mode of $g_s$.  It is replaced by 7-brane moduli, which we will discuss later.

This potential is minimized at
\beq\label{stabmod} R_f\sim \epsilon R_{AdS} \ ,~~~ R\sim \epsilon^{1/2} R_{AdS}\ ,
~~~ R_{AdS}^4\sim \frac{N_c}{\epsilon^3}\ , \eeq
with the desired hierarchy between the internal and $AdS$ directions.  It is interesting to note that since the internal volume satisfies $ Vol \sim R_{AdS} N_c^2$, this solution satisfies the weak gravity conjecture of \cite{ArkaniHamed:2006dz}\ (see the discussion around equation (13) of \cite{ArkaniHamed:2006dz}).

\section{$AdS_5\times Small$ Examples}
\label{sec:specific}

In this section we will describe our simplest examples implementing the strategy just outlined, developing a class of small-radius compactifications down to $AdS_5$.  Later in the paper, we will generalize to $AdS_4$.

In order to control the contributions of the 7-branes to the curvature and to the resulting potential energy
in five dimensions, we will use F theory \cite{Vafa:1996xn,MorrisonVafa}. This naturally incorporates the back reaction of the 7-branes in IIB, while geometrizing the problem.

In order to study the elliptically fibered geometry on which F theory is formulated, we will use the technique introduced in \cite{Witten:1993yc}, obtaining the geometry as the IR limit of the target space of a two-dimensional (2,2) supersymmetric gauged linear sigma model (GLSM).  In our case, the infrared limit of this two-dimensional sigma model is not a string worldsheet theory; we will just use the sigma model as a crutch for understanding the geometry and its symmetries.  Rather than review here the construction of GLSMs, we refer the unfamiliar reader to \cite{Witten:1993yc}\ for a clear introduction.

\subsection{Brane Construction}
\label{subsec:braneconstruction}

Let us first construct the noncompact brane systems whose infrared limit will give our field theories.  This will consist of $N_c$ D3-branes in F theory on a noncompact Calabi-Yau fourfold, preserving $d'=4, {\cal N}=1$ supersymmetry in the CFT.  In order to obtain a hierarchy $R_{\cal Y}\ll R_{AdS}$ (where now $R_{\cal Y}$ is the size of ${\cal Y}$ in its longest direction), we will choose our example as follows so that the cross sectional size $R_{\cal Y}$ is parametrically smaller than the radial distance from the origin.

To this end, consider a GLSM with chiral superfields $(\Phi_0,\Phi_1,\Phi_2,\dots,\Phi_D, X,Y,Z,P)$ with the following charges under a $U(1)^2\times U(1)^{D-3}$ gauge group:

\smallskip
\beq\label{chargevectors} \begin{tabular}{l l l l l l l l l}
  $\Phi_0$ &$~\Phi_1$ &$\Phi_2$ &$\dots$ &$\Phi_D$ &X &Y &Z &P\\
  \hline
                 0     & ~0     &0      &$\dots$ & 0     &2  &3&1 &-6 \\
                $-w_0$   & $~w_1$   &$w_2$    &$\dots$ & $w_D$   & 0 &0 &$w_0-\sum_{j=1}^D w_j$ & 0 \\
                $Q_0^a$  & $~Q^a_1$ &$Q^a_2$  &$\dots$ & $Q^a_D$ &0  &0 &$-\sum Q^a$ & 0\\
\end{tabular} \eeq

\bigskip

%
%

\noindent where $a=1,\dots,D-3$.  Here we take all the $w_I$, $I=0,\dots D$, to be positive (note the sign convention on $w_0$). The fact that the charges sum to zero means that the Calabi-Yau condition is satisfied for the target space geometry of the GLSM.

The 7-branes we are interested in -- equivalently the elliptic fiber of the Calabi-Yau fourfold -- are incorporated via a superpotential of the form
\beq\label{GLSMsup} \int d^2\theta P\left[Y^2-X^3-XZ^4f(\Phi_0,\Phi_1,\dots,\Phi_D) - Z^6 g(\Phi_0,\Phi_1,\dots,\Phi_D)\right]\ .
\eeq
Consider first the target space of this model at fixed values of the $\Phi_I,I=0,\dots, D$.  The first row of charges in (\ref{chargevectors}) combined with the superpotential (\ref{GLSMsup}) describes a $T^2$ realized as a surface in a weighted projective space $W\! P^{(2)}_{231}$.  The complex structure of this $T^2$ varies as a function of the coordinates $\phi_I$.  Gauge invariance under the second $U(1)$ above requires the degrees of $f$ and $g$ to be
\beq\label{GLSMdegrees} deg_f=4(\sum_{j=1}^Dw_j-w_0) ~~~~~ deg_g=6(\sum_{j=1}^Dw_j-w_0) \eeq
under the weighted identifications imposed by the charges $w_I$.

Singularities of the elliptic fiber -- places where the discriminant $\Delta=4f^3+27 g^2$ vanishes -- correspond to 7-branes.
The weights restrict the form of the superpotential (\ref{GLSMsup}).  We must ensure that the polynomials $f$ and $g$ can be chosen sufficiently generally so as to avoid disallowed singularities in the IR target space of the GLSM.  In general, it is not known which behaviors are allowed.  Because the GLSM respects (2,2) supersymmetry with non-anomalous $U(1)\times U(1)$ R symmetries which are consistent with the required R symmetries of an IR (2,2) superconformal field theory, it appears that the supersymmetry is generally preserved; the question then becomes one of whether the space decompactifies in the infrared.  At codimension one, for sufficiently high-order vanishing of $f$ and $g$ there are examples in which the target space does decompactify in the infrared.  In \S3.1\ below, we will suggest a sufficient condition for avoiding such a decompactification.

The scalar potential of the GLSM has, in addition to the F-terms generated by the superpotential (\ref{GLSMsup}), the D-terms
\begin{eqnarray}\label{GLSMD} && \left(2|x|^2+3|y|^2+|z|^2-6|p|^2\right)^2 +
\left(-w_0|\phi_0|^2+\sum_{j=1}^Dw_j|\phi_j|^2 +(w_0-\sum_{j=1}^3w_j)|z|^2\right)^2 \nonumber\\
&& \qquad\qquad \qquad\qquad  + \sum_{a=1}^{D-3}\left(\sum_jQ^a_j (|\phi_j|^2- |z|^2) \right)^2\ . \end{eqnarray}
Altogether, this construction produces a noncompact, elliptically fibered Calabi-Yau fourfold ${\cal C}$ as the IR target space of the model.  One can alternatively use F terms instead of some or all of the $D-3$ additional $U(1)$ gauge projections to reduce the target space to four complex dimensions.

We will be interested in the theory of D3-branes at the origin $\phi_0=\phi_j=0$.  If we impose that the total unweighted degree of the superpotential terms (\ref{GLSMsup}) is constant, then both the F and D terms in the GLSM scale uniformly as one approaches the origin and we expect the GLSM metric to flow to that of a cone.\footnote{The metric may flow to a cone even without this homogeneity assumption, but this requires a significant radiative correction to the classical D-term metric, and we would like to use this metric as a guide to the geometry.}

We have set the Fayet-Iliopoulos (FI) parameters in the D terms to zero.  As explained in \cite{Witten:1993yc}, the running of these couplings is proportional to the sum of the gauge charges, which vanishes here.  For the first $U(1)$, we consider vanishing FI parameter because in F-theory the elliptic fiber is taken to be of vanishing size.  For the second $U(1)$, we take vanishing FI parameter in order to obtain a conformal field theory from D3-branes at the tip $\phi_0=0=\phi_j$, $j=1,\dots 3$.  However, we will encounter some subtleties at the tip of the cone in our explicit examples below, which we will regulate by turning on FI parameters.


By choosing the weights $w_I$ appropriately, we can obtain a hierarchy between the internal and $AdS_5$ radii. A hierarchy will arise when the noncompact geometry determined by the above specifications takes the shape of a very narrow cone.
In order to check this, we need metric information.  The metric determined by the kinetic terms in the GLSM is not protected aside from holomorphic quantities; it flows to the Calabi-Yau metric in the infrared.  Explicit metrics on Calabi-Yau manifolds are difficult to obtain in general.
However, the ultraviolet metric in the GLSM gives a good estimate of the opening angles of the cone in examples where the exact metric is known, such as orbifolds and more general Calabi-Yau 3-fold cones.
We will therefore start by analyzing the UV GLSM metric, comparing cases with different choices of charges $w_j,Q^a_j$ to see which produce a narrow cone in this metric.  Then, we will show how this result agrees with a direct analysis of the near-horizon energetics, connecting the present construction to the stabilization mechanism described in \S2.
This final step provides concrete evidence for the usefulness of the GLSM metric as a guide to the shape of the cone.

Consider the regime
where the weights $w_j$,  $j=1,\dots D$, are all approximately equal to a large value $w\gg w_0$:
\beq\label{wregime} w_0 \ll w_j ~~~{\rm and} ~~~ w_i-w_j\ll w_j\approx w, ~~~~ i,j=1,\dots, D.\eeq
The kinetic terms in the UV regime of the GLSM are flat:
\beq\label{metricUV} ds^2=|d\phi_0|^2+\sum_j|d\phi_j|^2 \eeq
At a distance $|\phi_0|$ from the origin, the D-terms (\ref{GLSMD}) enforce that the remaining fields trace out a 5-manifold.  The second D-term in (\ref{GLSMD}) ensures that this is small in the $\phi_j$ directions in the regime (\ref{wregime}):
\beq\label{narrow} {\rm radius}^2\sim |\phi_0|^2\sim {w\over w_0}|\phi_j|^2 ~~~~~ j=1,\dots,D \ .\eeq

Let us now describe the geometry more precisely, keeping track of the angular directions in field space.
Writing $\phi_I\equiv \rho_Ie^{i\gamma_I}$, $I=0,1,\dots, D$, the GLSM kinetic terms take the form
\beq\label{GLSMkinetic} \int d^2\sigma \left(\sum_I\rho_I^2(\del\gamma_I+w_IA+Q^a_IB^a)^2 + \sum_I (\del\rho_I)^2\right) \ ,
\eeq
where $A$ is the gauge field corresponding to the second $U(1)$ in (\ref{chargevectors}) and $B^a$ are the gauge fields corresponding to the $U(1)^{D-3}$ gauge symmetry encoded in the last set of charge vectors in (\ref{chargevectors}).
Let us focus on the effects of integrating out the gauge field $A$, taking $D=3$ so there are no $B^a$ gauge fields.  Integrating out $A$ reduces the kinetic terms for the angles $\gamma_I$ to
\beq\label{kineticreduced} \int d^2\sigma \left(\rho_0^2 (\del\gamma_0)^2 + \sum_{j=1}^3\rho_j^2(\del\gamma_j)^2-
\frac{(-\rho_0^2 w_0  \del\gamma_0 + \sum_{j=1}^3\rho_j^2w_j\del\gamma_j)^2}{\rho_0^2w_0^2+\sum_j\rho_j^2w_j^2}  \right) \ .
\eeq
There is still a gauge redundancy which we could fix by setting $\gamma_0 = 0$, but it is convenient to keep it for now.
Recall that $\rho_0^2\sim (w/w_0)\rho_j^2$, Eq.~(\ref{narrow}).  Consider first a circle where {one} of the $\gamma_{i}$ goes from zero to $2\pi$.  The second and third terms in the metric are comparable but do not cancel, and the radius of this circle is of order $\rho_i$.  This is the same scale
\begin{equation}
\rho_0 \sqrt{w_0/w}
\label{size}
\end{equation}
as found above.  However, now consider the circle where $\gamma_1 = \gamma_2 = \gamma_3 = \theta$.  The second and third terms now cancel up to remainders of relative order $w_0/w$, $(w_i - w_j)/w$, and so this circle is parametrically smaller by a $1/\sqrt w$.   Identifying this circle with the fiber circle $S^1_f$,
altogether we have the relations $R^2\equiv |\phi_j|^2\sim |\phi_0|^2/w$ and $R_f\sim |\phi_0|/w$.
These features are just as in the geometry (\ref{hopf}) discussed in \S2.  So our cone, according to the GLSM metric, has a hierarchy of the form (\ref{stabmod}) with $\epsilon\sim w_0/w$.

Finally, consider the circle $\gamma_{0,1,2,3} = \theta$, which is the phase conjugate to the overall rescaling of the cone.  The final term in the metric vanishes identically, and the first term dominates.  The radius is thus order $\rho_0$, which is no smaller than the distance from the tip.  We might have expected this: this configuration of 3-branes and 7-branes leaves $d=4$, ${\cal N}=1$ supersymmetry, for which the $U(1)$ $R$-symmetry often protects such a large circle as discussed in Sec.~2.1.  However, this circle is much larger than the actual radius of the compact space: it is actually wound multiple ($w/w_0$) times around the fiber direction.  To see this, first consider the gauge-equivalent circle
\begin{equation}
\gamma_0 = 0, \gamma_i = (1+w_i/w_0)\theta\ .
\label{reeb}
\end{equation}
  For $\theta = 2\pi w_0/w$, all angles are $2\pi + O(1/w)$, and so the distance traveled is only of order $w^{-1} \cdot w^{-1/2} = w^{-3/2}$, where the factor $w^{-1/2}$ is from Eq.~(\ref{size}): we have gone a distance $O(1/w)$ around the fiber and ended up close to our starting point.
Correspondingly this implies, as expected, that the lightest KK states are characterized by the overall radius $\rho_i \sim \rho_0 \sqrt{w_0/w}$ and not the larger radius of this circle.  If there is a gradient along the large circle (i.e. an $R$-charge), then there is a much larger gradient in the orthogonal directions.

\subsection{Near Horizon Compactification Geometry}

Let us derive this again in a second way, directly in the near horizon geometry.  As we have just seen, the base ${\cal B}$ in (\ref{hopf}) is given by the geometry at fixed $\phi_0$, and the circle fiber is the $U(1)$ direction with charges $w_I$.  Let us formulate F-theory on this space therefore using the
above GLSM without the field $\phi_0$.  Without $\phi_0$, the sum of the charges of the second $U(1)$ does not cancel, so this theory has a running Fayet-Iliopoulos parameter $R^2$ for this $U(1)$; that is, the D-terms now take the form
\begin{eqnarray}\label{GLSMNHD} & \left(2|x|^2+3|y|^2+|z|^2-6|p|^2\right)^2 +
\left(-R^2+\sum_{j=1}^Dw_j|\phi_j|^2 +(w_0-\sum_{j=1}^3w_j)|z|^2\right)^2 \\
&+ \sum_{a=1}^{D-3}\left(\sum_jQ^a_j|\phi_j|^2-(\sum_j Q^a_j)|z|^2\right)^2\ . \end{eqnarray}

The running of $R^2$ is given by the sum of the gauge charges:
\beq\label{betafull} \beta_{R^2}^{{\cal B}+7Bs}\sim \sum_j w_j-(\sum_jw_j-w_0)=w_0\ . \eeq
Here we have separated this into the contributions from the $\phi_j, j=1,\dots,3$ and the contribution from $z$.  The latter contribution has to do with the 7-branes.
The net beta function (\ref{betafull}) is parameterically smaller than it would be in the absence of the 7-branes:
\beq\label{betanoseven} \beta_{R^2}^{{\cal B}}\sim \sum_jw_j\ . \eeq
This implies that in \ref{stabmech}, the small parameter $\epsilon$ is given by
\beq\label{epsweighted} \epsilon\sim \frac{\beta_{R^2}^{B+7Bs}}{\beta_{R^2}^{B}}\sim \frac{w_0}{w}
\ .\eeq
That is, our setup ensures that the 7-branes nearly cancel the positive curvature energy of ${\cal B}$, realizing our original strategy outlined in \S2.

\subsubsection{7-Brane Moduli and Dilaton}

In our discussion of the geometry and stabilization mechanism thus far, we have suppressed the dependence on the dilaton.  In general, the type IIB dilaton varies as a function of position in F theory models.  In general, the moduli of the 7-branes are encoded in the complex moduli of the elliptically fibered manifold on which F theory is formulated.  These appear in the superpotential in the GLSM formulation of the space. If we start at an enhanced symmetry point (where the 7-branes realize unbroken gauge symmetry in the bulk), the system is at an extremum of the full quantum effective potential.
Superpotential couplings are protected from perturbative renormalization.  These directions are therefore flat to all orders in perturbation theory.  As such, even if the enhanced symmetry points we consider turn out to be maxima rather than minima, they correspond at worst to BF-allowed tachyons~\cite{Breitenlohner:1982jf} in $AdS_5$, because of the supersymmetry of the solution.  In particular, this means that we expect no disallowed tachyonic modes where 7-branes slip off of contractible cycles in ${\cal Y}_5$.

\subsection{A Criterion for Allowed Singularities}
\label{sec:singularities}

Because of the possibility of singularities in the physics, not all models in the class just outlined will be consistent.  Some geometrical singularities are physically allowed in F theory -- such as those corresponding to nonabelian gauge symmetry on the 7-branes -- and some are not.  It is not generally known which is which.  In this subsection we describe a criterion for allowed singularities.

In particular, we need to determine the conditions under which no decompactification limits arise in our compactification geometry.  We will start by analyzing in the context of perturbative string theory on our noncompact $CY_4$.   In that case the infrared limit of the GLSM describes the worldsheet of a string.  There we have methods to analyze singularities, combining the tools developed in \cite{Witten:1993jg}\ and \cite{Silverstein:1995re}\ using the GLSM framework \cite{Witten:1993yc}.  Although it will be derived in the context of perturbative string theory, our criterion will coincide with the known conditions on singularities involving coincident 7-branes on a $CP^1$.  This criterion would apply directly in type IIA string theory, which is dual to F theory compactified on an additional $T^2$.  We will make further comments on the application to F theory below.

In the GLSM, singularities arise in the worldsheet path integral from regions in field space where scalar fields can go off to infinity.  When the polynomials defining the target space manifold are transverse, and the FI parameters and theta angles take generic values, this does not occur \cite{Witten:1993yc}.  In the class of models we outlined in the previous section, the weights $w_I$ in general restrict the form of the polynomials, leading to examples in which they are non-transverse.  When the polynomials are non-transverse, the scalar potential of the GLSM no longer forces $p$ to vanish, and there is a branch in which $p$ goes to infinity along with $z$ and some subset of the $\phi_I$'s, constrained by the condition that the GLSM D-terms vanish.    This defines a noncompact branch in field space of some dimension $d_{sing}$.  Naively one might think that this constitutes a disallowed decompactification limit.  However, the situation is more nuanced than that -- after all ALE singularities and the conifold singularity are both examples of this phenomenon \cite{Silverstein:1995re}, and although singular at the level of the worldsheet theory, the spacetime theories in these cases are benign (involving an additional finite set of light fields).

In these previously understood cases, the singularity is equivalent to a linear dilaton throat.  A simple way to see this is that the central charge along these directions is less than it is in bulk, on the branch where $p=0$ and the $\phi_j$ trace out the Calabi-Yau geometry.  To match the throat onto the bulk, a spacelike linear dilaton makes up the difference in central charge.  This produces a gap in the spectrum of string states, explaining the absence of a truly singular tower of Kaluza-Klein modes, as one would have in a decompactification limit.  In particular, compactification on spaces including such throats still leads to a finite Planck mass in the remaining dimensions.

This suggests a rather simple criterion:  a singularity is allowed if the central charge $\hat c_{throat}$ in the throat is less than that in bulk ($\hat c_{bulk}=4$ in our case of Calabi-Yau fourfolds)
\beq\label{singcond} \hat c_{throat} < 4\ .\eeq
In calculating
$\hat c_{throat}$, it is crucial to include not just $d_{sing}$ defined above, but also contributions from all fields in the throat, even those that do not have a flat direction in their potential, as long as they are massless.  The contribution of massless fields to $\hat c$ in the GLSM was developed in \cite{Witten:1993jg}.  It depends on the degrees with which the massless fields appear in the GLSM superpotential; higher degrees lead to larger contributions to $\hat c$.

Let us start by analyzing this in the well understood case of K3 realized as an elliptic fibration over $CP^1$.
This is described by fields $(\Phi_1,\Phi_2,X,Y, Z, P)$ with charge vectors $(0,0,2,3,1,-6)$ and $(1,1,0,0,-2,0)$
under a $U(1)\times U(1)$ gauge symmetry.  The polynomial $g(\phi_1,\phi_2)$ appearing in the superpotential (\ref{GLSMsup}) is of degree 12, and $f(\phi_1,\phi_2)$ is of degree 8, leading to the presence of 24 7-branes at the points where $\Delta=27g^2+4f^3=0$.   Consider a point where $g$ vanishes at $\phi_1=0$ with degree $n$ ($g\sim \phi_1^n \phi_2^{12-n}$) with $f$ vanishing with degree $n_f\ge 2n/3$.  At $\phi_1=0$, the superpotential does not constrain $\phi_2$, and there is a branch in scalar field space where $z,p,$ and $\phi_2$ go off to infinity constrained by the two D-terms, giving $d_{sing}=1$.  Along this branch, the GLSM superpotential for the other fields is of the form
\beq\label{Wother} W_{sing} = Y^2-X^3-\langle Z\rangle^6\Phi_1^n - X \langle Z\rangle^4 \Phi_1^{n_f} \eeq
Along this branch, $Y$ is massive, but $X$ and $\Phi_1$ are massless.  As explained in \cite{Witten:1993jg}, the fields in the superpotential contribute central charge $\hat c=\sum_i (1-2\alpha_i)$ where for a quasihomogeneous superpotential the $\alpha_i$ are related to the degree $I_i$ of $W$ in the various chiral superfields $\eta_i$ via the relation
\beq\label{quasihom}\sum_i \alpha_i\eta_i\del_i W=W \Rightarrow \sum_i\alpha_iI_i=1 \eeq
Intuitively, the central charge is reduced from the free field value by an amount which goes inversely with the degree of the superpotential.  We can consider for simplicity $f=0$; then $X$ and $\Phi_1$ do not mix.  This gives us $\alpha_X=1/3$ (so $X$ contributes 1/3 to $\hat c$), and $\alpha_{\Phi_1}=1/n$ (so $\Phi_1$ contributes $1-2/n$).  The total $\hat c$ along the throat is therefore $\hat c_{throat}=d_{sing}+1/3+1-2/n$.  In order for this to not decompactify by our above criterion, we require $\hat c_{throat}<2\Rightarrow n<6 $.

In particular, for $n = 6$, although in the UV GLSM metric the point $\phi_1=0$ lies at finite distance, the GLSM kinetic terms get renormalized in the IR to give the metric of flat $T^2\times S^1\times R$; the K3 has decompactified. In the F theory language, this is precisely the standard criterion to avoid introducing so many 7-branes that they source a $2\pi$ deficit angle, causing such strong back reaction that the base $P^1$ decompactifies to becomes an infinite cylinder.

We will impose (\ref{singcond}) more generally on our compactifications.  If this criterion is not satisfied, so that there is a noncompact throat with $\hat c\ge 4$, we expect that the five dimensional Planck mass is infinite since there is no linear dilaton down the throat and the graviton wavefunction is not massed up.  In examples in the next subsection we will encounter a marginal singularity -- one with $\hat c_{throat}=4$ -- near the tip of the cone, and will analyze this separately.

Let us now apply this criterion to our class of examples, and explain some simple would-be examples which are eliminated by our criterion.  Consider the case $D=3$ with weights $-w_0,w-\delta, w, w+\delta$ for the fields $\phi_I$, $I=0,\dots,4$, with $w\gg w_0,\delta$.  This class of models would give a compactification on a hopf fibration over the weighted projective space $W\!P^2_{w-\delta,w,w+\delta}$.  In order to satisfy the degree condition (\ref{GLSMdegrees}), the polynomials $f$ and $g$ in (\ref{GLSMsup}) are significantly constrained.  Consider for example a simple set of models where $\phi_0$ does not appear in the superpotential.  Then the polynomial $g$ is
\beq\label{WPfg} g\sim \sum_{I=3w_0/\delta}^9 \phi_1^{I+3w_0/\delta}\phi_2^{18-2I}\phi_3^{I-3w_0/\delta} \eeq
and the polynomial $f$ behaves analogously.
This model has a singular branch on which $p,z$, and $\phi_3$ blow up together, consistently with the vanishing of the scalar potential, with $\phi_1=\phi_2=x=y=0$.  On this branch, $p,z,$ and $\phi_3$ together carry one unit of $\hat c$, $\phi_0$ contributes one unit, and
$x$ carries central charge $\hat c_X=1/3$.   Imposing (\ref{quasihom}), we obtain $\alpha_1=2\alpha_2$ and $\alpha_2=1/(18+6w_0/\delta)$.  This implies that $\phi_1$ and $\phi_2$ carry $2-1/(3+w_0/\delta)$ units of $\hat c$.  Altogether, this branch carries $\hat c_{throat}=1+1+1/3+2-1/(3+w_0/\delta)>4$.  Because this is greater than $\hat c_{bulk}=4$, the model is singular.

\subsection{Some Examples}

However, we can generalize the construction slightly to obtain an infinite sequence of nonsingular examples.
Consider a GLSM with charges under a $U(1)^3$ gauge group corresponding to the following D-terms
\begin{eqnarray}\label{modelD} & \left(-2|\phi_0|^2+(w+1)|\phi_1|^2+w(|\phi_2|^2+|\phi_3|^2)-(3w-1)|z|^2 - r_1\right)^2 \\ \nonumber
& + (2|\eta|^2-|z|^2+{1\over 3}|\phi_1|^2-{2\over 3}(|\phi_2|^2+|\phi_3|^2)-r_2)^2 + (2|x|^2+3|y|^2+|z|^2-6|p|^2)^2
\end{eqnarray}
We have allowed for nonzero Fayet-Iliopoulos terms.  To begin with, let us set these to zero; we will later use them to analyze the theory near the tip of the cone ${\cal C}$ swept out by the fields $\phi_I,\eta$.

In this model, the term in the superpotential involving the polynomial $g$ (satisfying gauge invariance and homogeneity in unweighted rescalings of the fields) is
\beq\label{modelW} \int d^2\theta P z^6\left( \sum_{a,I}\eta^{9-a/2}\phi_0^{3+a/2}\phi_1^{a}\phi_2^I\phi_3^{18-a-I}\right) \eeq
and there is a similar expression for the polynomial $f$.
This has total unweighted degree 30, uniformly in all terms.  As above, in order to check for singularities we must analyze the infinite branches in scalar field space which arise in this model.  Again these branches arise when $p$ and $z$ grow large together.  In the present model, this also implies that $\eta$ blows up as we can see as follows using (\ref{modelD}).
Solving the first D-term for $|z|^2$ and plugging into the second yields
\beq\label{Dagain} \left(2|\eta|^2-({2\over 3}+{w\over{3w-1}})(|\phi_2|^2+|\phi_3|^2)-{4\over{3(3w-1)}}|\phi_1|^2+{2\over{3w-1}}|\phi_0|^2\right)^2 \eeq
Combining this with the first term in (\ref{modelD}), which requires $\phi_j$ to blow up for some $j=1,2,$ or $3$, shows that $\eta$ must diverge on any singular branch.
So the question of singularities is reduced to the analysis of the regimes where two or three of the fields $\phi_0,\phi_1,\phi_2,$ and $\phi_3$ vanish while $p,z,\eta,$ and at least one of the $\phi_j, j=1,2,3$ diverge.

This model, and many others like it that we have analyzed, has a marginal singularity, but only one emanating from the tip of the cone at the origin of field space (along a branch ${\cal S}$ where $\phi_0,\phi_2,$ and $\phi_3$ vanish and where $p\propto z\propto\phi_1\propto\eta$ turn on).  This is a ``hybrid" space in GLSM terminology:  part of the central charge arises from large, geometric dimensions and part from a string-scale Landau-Ginzburg theory transverse to these dimensions. At the point $\phi_0=\phi_1=\phi_2=\phi_3=\eta=0$, there are actually several branches which join together:  the cone ${\cal C}$ of interest at nonzero $\phi_0$, the branch ${\cal S}$ we just mentioned, and ``$\sigma"$ branches in which the adjoint scalars $\sigma_\alpha$ of the GLSM turn on.

We would like to understand if the GLSM metric renormalizes strongly enough to decompactify the tip of the cone, sending it off to infinite distance.  In order to analyze this, let us first regulate the problem by turning on a negative FI parameter $r_2$ in (\ref{modelD}).  In our (2,2) supersymmetric system, $r_2$ is part of a complex parameter $t_2=r_2+i\theta_2$, pairing up with the theta angle $\theta_2$ of the second $U(1)$ in (\ref{modelD}) \cite{Witten:1993yc}.  In the application of this sigma model to type II string theory on the Calabi-Yau fourfold ${\cal C}$, $t_2$ corresponds to a complex scalar modulus field in spacetime, part of a chiral multiplet; the spacetime superpotential depends on it holomorphically.  More abstractly, one can define a topologically twisted sigma model whose observables all vary holomorphically with $t_2$.
In this system with $t_2\ne 0$, with $r_2<0$, the fields $\phi_2$ and $\phi_3$ cannot both vanish, and so we have disconnected the branch ${\cal S}$ from our Calabi-Yau target space ${\cal C}$.  We would like to understand if physical correlation functions behave as if the tip decompactifies in the limit $r_2\to 0$.  Let us probe this question with holomorphic quantities; since these exist down the supersymmetric branch ${\cal S}$, we expect that they are sufficient to detect decompactification.  These quantities can only become singular at a special value of the holomorphic parameter $t_2$.  So if we keep $\theta_2$ generic, the system is nonsingular at the tip, at least as probed by holomorphic quantities.

As described in the previous section, there is a large zoo of potential examples.  With a large number of independent fields $\phi_j$, $j=1,\dots D$ it may be possible to obtain examples without marginal singularities at the tip.  It would be very interesting to analyze this systematically.

\subsection{Entropy}
\label{sec:entropy}

The brane construction we have developed matches the near horizon stabilization mechanism described in \S2.  It is still a complicated problem to derive the field theory from the low energy limit of this brane construction.  However, it is straightforward to determine the leading parametric dependence of the number of degrees of freedom of the CFT on our discrete parameters and to obtain a heuristic interpretation of this number, as follows.  In general,
\beq\label{Ndofgeneral} N_{d.o.f.} \sim M_5^3 L_{AdS}^3 \eeq
where $L_{AdS}\equiv R_{AdS}\sqrt{\alpha'}$ is the $AdS_5$ radius and $M_5$ is the five-dimensional Planck mass.
From the scaling (\ref{stabmod}) we find
\beq\label{Ndofus} N_{d.o.f.}\sim \frac{N_c^2}{\epsilon^3}\sim w^3N_c^2 \eeq
Recall that the hierarchy of length scales (\ref{stabmod}) in our solution, which leads to the result (\ref{Ndofus}), is tied to the narrowness of the noncompact cone defining our brane construction as discussed in \S3.1.  Consider a few probe D3-branes pulled away from the tip of the cone by a distance $L_{rad}$.  This corresponds to our field theory out on its (approximate) moduli space.  There are degrees of freedom in this theory given by strings which stretch between the D3-branes.  Because the cone is narrow, a string which stretches around the cone a distance $L_{cone}$ (of mass $L_{cone}/\alpha'$) is lighter than one which stretches radially to the tip (of mass $L_{rad}/\alpha'$).  Similarly, in the near horizon region, a string stretching around the compactification  is lighter than one stretching down to the AdS horizon.  In fact (expressing the sizes in string units) there are of order
\beq\label{woundstrings} N_{wound}\sim \left(\frac{R_{rad}}{R_f}\right)\left(\frac{R_{rad}}{R}\right)^4
\sim \frac{1}{\epsilon^3}   \eeq
wound strings which are lighter than a single string extending to the tip.  This agrees with the parametric dependence in (\ref{Ndofus}).

This is similar to the situation in e.g. $Z_k$ orbifold conformal field theories \cite{Kachru:1998ys}, where the wound strings correspond to bifundamental matter which builds up an entropy of order $k N_c^2$.  The difference in the present case is that all internal directions are parameterically smaller than the radial distance to the tip of the cone.  In the orbifold case, strings stretching to the tip can unwind, so the estimate analogous to (\ref{woundstrings}) saturates the entropy.  In the present case, the estimate (\ref{woundstrings}) is a lower bound on the entropy.  Although the tip region is complicated, it is tempting to conjecture that strings can unwind there in the present case as well.

Another argument pointing to the same conclusion is to look at the complex base $\cal B$ obtained by fixing the radial coordinate and modding out the phase~(\ref{reeb}).  The base has a $Z_{w_i/w_0}$ singularities when the two coordinates $\phi_j$, $j \neq i$, vanish.  These are not supersymmetric, but do not lead to tachyons because the full space with the phase direction included is smooth.  However, they suggest that a $Z_{w_1/w_0} \times Z_{w_2/w_0} \times Z_{w_3/w_0}$ quiver may be present.

\section{Further Directions}

We have obtained a class of brane constructions whose low energy limits give field theories dual to small-radius compactifications.  This is motivated by the basic goal of formulating four dimensional quantum gravity in string theory, as well as the goal of developing new corners of the landscape amenable to simple and controlled model building.\footnote{It is interesting to consider applications both to models of particle physics and cosmology, and to theoretical states of matter as in \cite{Denef:2009tp}.}

In this final section, we will first describe an immediate generalization of our five-dimensional construction above to the four-dimensional case of most physical interest.  Next we will comment further on the field theory duals.  We will then explain potential generalizations in which orientifolds provide the negative potential energy, suggesting a concrete generalization to de Sitter minima.  Finally we comment on the prospects of connecting this work to the problem of formulating four-dimensional cosmology non-perturbatively.

\subsection{$AdS_4\times Small$ Generalizations}
\label{subsec:AdS4}

In the previous sections, we focused on a relatively simple set of compactifications down to $AdS_5$.  We can generalize this to $AdS_4$ in two ways.
The first method for reducing from what we have done to four dimensions is to study M2 branes in M theory at the tip of the  Calabi-Yau fourfold cone that we constructed.  This gives a hierarchy in terms of pure geometry, with the elliptic fiber part of space in M theory (as opposed to F theory where it describes the axio-dilaton).

The second method to get down to four dimensions is to tensor in another circle, considering $S^1\times {\cal Y}_5$.  First, to warm up consider adding 1-form flux along the $S^1$.  This stabilizes it at a large radius of order $R_{AdS}\sqrt{\alpha'}$, as follows.
The potential is of the form (with radii given in string units and $g_s\sim 1$)
\beq\label{Umod} {\cal U}\sim M_P^4 (R_f R_6 R^4)^{-1}\left(\frac{R_f^2}{R^4}-\frac{\epsilon}{ R^2}+\frac{N_c^2}{R^8R_f^2}+\frac{Q_1^2}{R_6^2}\right)  \eeq
where $Q_1$ is the 1-form flux quantum number along the new $S^1$ of radius $R_6\sqrt{\alpha'}$.
Extremizing the potential
with respect to $R$, $R_f$, and $R_6$, we get a solution with
\beq\label{scalingsfour} R_f^2\sim \epsilon R^2, ~~~~~ R^4\sim \frac{N_c}{\epsilon}, ~~~~~
R_{AdS}^2\sim \frac{R^2}{\epsilon}\sim \frac{R_6^2}{Q_1^2} \ . \eeq

From this we see that one unit of one-form flux leads to $R_6 \sim R_{AdS}$, not giving a full hierarchy.  However, we can obtain a hierarchy with $R_6\ll R_{AdS}$ if we consider instead 3-form flux along the new $S^1$ times a 2-cycle in ${\cal Y}_5$, since the 3-form flux is parametrically more dilute than one-form flux.  Replacing $Q_1$ in (\ref{scalingsfour}) with $Q_3/R^2$ leads to a solution with $R_6^2\sim Q_3^2/(R^2\epsilon)$.
In the zoo of examples outlined above in \S3, many have a rich topology with one or more 3-cycles and dual 2-cycles \cite{Ypq,Labc,Hananyetal}\ on which to put this 3-form flux.

\subsection{The CFT Duals}

The field theories dual to our hierarchical models are defined indirectly by the low energy limit of our brane construction.  We would like a more direct presentation of their content and couplings.
The discussion in section~\ref{sec:entropy}\ gives some clues as to the nature of the field theories, but this is far from a complete characterization analogous to that available for toric Calabi-Yau three-fold cones \cite{Hananyetal}.
Had we needed only D3- and D7-branes, we could presumably determine its content by moving the D7-branes away, finding the quiver gauge theory for the toric D3 theory \cite{Hananyetal}, and then adding appropriate fundamental matter.  With $(p,q)$ 7-branes we need to add mutually nonlocal dyonic fields.  Thus we have a fixed point of Argyres-Douglas type~\cite{Argyres:1995jj}, for which one cannot directly write down a Lagrangian~\cite{JP}.

However, one can likely flow to such a theory starting from a purely electric theory in the \mbox{UV}.
In the present case one way to try to identify this theory would be to choose the 7-brane moduli to lie at an orientifold point, where the polynomials $f$ and $g$ are of the form $f \propto h^2$, $g \propto h^3$~\cite{Sen:1996vd}.  The CFT is then determined as an orientifold of one without any 7-branes.   However, in this limit there is an additional coordinate $\xi$ along with an embedding condition $\xi^2 = h$, requiring a superpotential in the GLSM, and so we are still not in the toric case~\cite{Hananyetal} where the duality is best understood.  Also as mentioned in section \ref{sec:specific}\ a subclass of examples employ additional F terms to define the target space geometry, which similarly takes us out of the class of purely toric constructions.

We hope that our work encourages the development of power tools to deal with non-toric spaces and 7-branes in AdS/CFT duality. Similar comments apply to the M theory examples of section \ref{subsec:AdS4}\ appropriate to the $(AdS_4\times Small)/CFT_3$ duality; one requires generalizations of \cite{ABJM}\ which apply to these non-toric geometries.

The appearance on the field theory side of electric and magnetic flavors is a direct consequence of our mechanism for lifting the curvature energy on the gravity side to obtain a hierarchy.  The significance of this relation between
four dimensional quantum gravity in string theory and Argyres-Douglas type field theories deserves further reflection.\footnote{It would be interesting to look for a hierarchy in other systems with rich contributions to the curvature energy, such as that recently discussed in \cite{Gaiotto:2009gz}.  So far these are non-hierarchical, but it may be interesting to seek new contributions or special limits where a hierarchy arises starting from these models, for which a detailed dictionary is already known.}

\subsection{Potential Generalizations and Cosmological Holography}

Another way to obtain a hierarchy of scales is to use 7-branes to fully cancel the curvature potential energy.  This removes the original negative term in the potential entirely.  In order to stabilize moduli, it is crucial to have sufficiently strong negative terms in the potential \cite{Landscapereviews}\ since all sources of potential energy dilute at large radius and weak coupling.  Negative terms can arise from orientifolds for example.  It would be interesting to construct examples of this kind.

If this method also works, it suggests a method for generalizing to obtain de Sitter constructions.  This would proceed by slightly {\it over}-canceling the curvature energy, rather than under-canceling it, and obtaining the negative term from orientifolds.  In this case, it would be very interesting to explore how the brane construction changes as we build up from AdS/CFT in such a way that the gravity side becomes a metastable de Sitter solution.  For further development of this idea see Ref.~\cite{Dong:2010pm}.

\section*{Acknowledgments}
We thank N. Arkani-Hamed, P. Aspinwall, M. Douglas, S. Franco, A. Hanany, S. Kachru, J. Maldacena, D. Martelli, D. Morrison, S. Shenker, J. Sparks, D. Tong, G. Torroba, S. Trivedi, and C. Vafa for useful discussions. E.S. thanks O. Aharony and D. Tong for many interesting discussions of this problem over the years, and S. Dimopoulos, S. Kachru, R. Kallosh, A. Linde, S. Shenker, and L. Susskind for many discussions of the landscape and its applications including the problem of its duals.  The research of
E.S. was supported by NSF grant PHY-0244728 and by the DOE under contract DE-AC03-76SF00515.
The research of J.P. is supported by NSF grants PHY05-51164 and PHY07-57035.

\begingroup\raggedright\endgroup

\end{document}